\documentclass[aps,prd,12pt,showpacs,notitlepage,nofootinbib,tightenlines]{revtex4-1}
\usepackage{amsmath}
\usepackage{bm}
\usepackage{times}
\usepackage{braket}
\usepackage{color}
\usepackage{epsfig}
\usepackage{slashed}
\usepackage{hyperref}
\usepackage{enumerate}
\usepackage{multirow}
\newcommand{\beq}{\begin{eqnarray}}
\newcommand{\eeq}{\end{eqnarray}}

\def \cpc{ {\bf Chin. Phys. C} }

\def \mpla{ {\bf Mod. Phys. Lett. A}  }
\def \copc{ {\bf Comput. Phys. Commum. } }
\def \epjc{{\bf Eur. Phys. J. C} }
\def \jpg{ {\bf J. Phys. G} }
\def \jcap{ {\bf JCAP}  }
\def \npb{ {\bf Nucl. Phys. B} }
\def \plb{ {\bf Phys. Lett. B} }

\def \prt{  {\bf Phys. Rept.} }

\def \prd{ {\bf Phys. Rev. D} }
\def \prl{ {\bf Phys. Rev. Lett.}  }

\def \zpc{ {\bf Z. Phys. C}  }
\def \jhep{ {\bf JHEP}  }

\definecolor{Red}{rgb}{1.,0.,0.}

\definecolor{Blue}{rgb}{0.,0.,1.}

\definecolor{nicered}{rgb}{0.7,0.1,0.1}
\definecolor{nicegreen}{rgb}{0.1,0.5,0.1}
\def\lsim{ {\ \lower-1.2pt\vbox{\hbox{\rlap{$<$}\lower6pt\vbox{\hbox{$\sim$}}}}\ } }
\def\gsim{ {\ \lower-1.2pt\vbox{\hbox{\rlap{$>$}\lower6pt\vbox{\hbox{$\sim$}}}}\ } }

\hypersetup{colorlinks,citecolor=nicegreen,linkcolor=nicered}

\begin{document}
\title{Search for single production of vector-like top partners at the Large Hadron Electron Collider}
\author{Yao-Bei Liu\footnote{E-mail: liuyaobei@sina.com}}
\affiliation{Henan Institute of Science and Technology, Xinxiang 453003, P.R.China }
\begin{abstract}
The new vector-like top partners with charge $2/3$ are a typical feature of many new physics models beyond the Standard Model (SM). We propose a search strategy for single production of top partners $T$ focusing on both the $T\to Wb$ and $T\to th$ decay channels at the Large Hadron Electron Collider (LHeC). Our analysis is based on a simplified model in which the top partner is an $SU(2)$ singlet, with couplings only to the third generation of SM quarks. We study the observability of the single $T$ through the processes $e^{+}p \to T(\to bW^{+})\bar{\nu}_{e}\to b\ell^{+}+ \slashed E_T^{miss}$ and $e^{+}p \to T(\to th)\bar{\nu}_{e}\to t(\to jj'b)h(\to b\bar{b}) \slashed E_T^{miss}$ at the LHeC with the proposed 140 GeV electron beam (with $80\%$ polarization) and 7 TeV proton beam. For three typical $T$-quark masses (800, 900 and 1000 GeV), the $3\sigma$ exclusion limits on the $TWb$ coupling are respectively presented for various values of the integrated luminosity.
\end{abstract}


\maketitle

\section{Introduction}
With the discovery of the 125 GeV Higgs boson at the Large Hadron Collider (LHC)~\cite{lhc-higgs}, our understanding of electroweak symmetry breaking~(EWSB) has been
significantly enhanced. However, some unanswered questions remain which forces us to look for new physics~(NP) beyond the Standard Model~(SM). One of the intriguing issues that needs
immediate attention is the naturalness problem~\cite{hmass}. Many popular models have been proposed and different solutions can be categorized based on the objects which cancel the largest Higgs radiative corrections
from the top quark. Some of these models postulate the existence of new heavy
fermions, such as little Higgs models~\cite{littlehiggs}, extra dimensions~\cite{ED} and composite Higgs models~\cite{twinhiggs,Agashe:2004rs}.
In many cases, these new fermions are vector-like top partners, whose common feature is to decay into a SM quark and a gauge boson, or a Higgs boson.
Many phenomenology studies for these new fermions have been made in the literature, see for example~\cite{p-lh,p-cp,p-twin,p1,p2,p3,NP,single,jhep1304-004,NP-shu,jhep1501-088}.

Here we mainly focus on the $SU(2)$ singlet $T$-quarks with charge $2/3$. Due to the
Goldstone-boson equivalence theorem, the branching ratios of $T$ into $bW$, $tZ$ and $th$ have a good approximation 2:1:1 in the limit $m_T\rightarrow \infty$. Previous studies of ATLAS and CMS Collaborations, relying on signatures induced by both
the vector-like $T$-quark pair-production and single-production
modes, impose strong constraints on the masses of the
heavy quarks that are now bounded to be above about
550-950~GeV~\cite{atlas-1,cms-1,atlas-cms-13}, depending on the assumed branching ratios. For the $Wb$ channel, the current constraint from the ATLAS Collaboration set a upper limit on the parameter $C^{Wb}_{L}<0.45$ which performed the search for singly produced vector-like $T$ quark at 13 TeV with 3.2 fb$^{-1}$~\cite{atlas13}.
Besides, the top partner couplings to
the SM particles are rather severely constrained by the electroweak precision observables~\cite{ew,rb}, various SM-like Higgs decay channels~\cite{higgs-decay} and the other indirect constraints~(see for example Refs.~\cite{Buchkremer:2013bha,bound}).
On the other hand, the crucial point is that even a small mixing to the first generation may have a
severe impact on single $T$-quark production processes~\cite{jhep2015-02-032,jhep1108-080}. In the previous studies~\cite{single,NP-shu,jhep1501-088}, it has been recognized that the single production of top partners at the LHC will provide the most promising channel in searching for a heavy top partner. In fact, the collider search for top partners has become and will remain an important
constraint on wide classes of new physics models. Thus it is
highly motivated to investigate all sensitive search
strategies within the possibly available accelerator and
detector designs.

In our present paper, we study the observability of a single $T$-quark production at the proposed Large Hadron-Electron Collider (LHeC). The LHeC~\cite{LHeC} would be the next high-energy $e$-$p$ collider which is designed to collide an electron beam with
a typical energy range, 60-150 GeV with a 7 TeV or higher proton
beam from the LHC. Its luminosity is projected to be as high as possibly 1 ab$^{-1}$, with a default value of 100 fb$^{-1}$~\cite{LHeC1}. Furthermore, the electron beam can be polarized and has an enormous scope to probe electroweak and Higgs boson
physics~\cite{LHeC2,LHeC3}. Certainly, it is possible that the new vector-like $T$-quarks can mix sizably with the SM light quarks and their
production cross section will be very large due to the mixing with valence quarks, but then their masses are not connected to EWSB and thus we do not consider this case. Although the $T\to tZ(\to \ell^{+}\ell^{-})$ channel is a primary option for most experimental searches, it has small number of events at the LHeC even for a high luminosity~\cite{plb-liu}. We expect that other channels will give a better constraint on the parameters of our model than the considered $T\to tZ$ channel. In this paper, we mainly study the observability of a single $T$-quark production at LHeC combine both
the $bW$ and $th$ decay channels for three typical masses of top partners.

This paper is organized as follows. In Sec.~II, we give a brief description of the simplified model including the vector-like $T$-quark with charge $2/3$. In Sec.~III we study the prospects of observing the single $T$ production by performing a detailed analysis of the signal and backgrounds in each channel. Finally, we draw our conclusions in Sec.~IV.

\section{Top partner in a simplified model}
It is clear that many extensions of the SM contain vector-like quark partners (and in particular top-partners), which can be classified according to their $SU(2)\times U(1)$ quantum numbers, where at least one of the partners needs to have the same electro-magnetic charge $2/3$ as the corresponding SM
quark. A generic parametrization of an effective Lagrangian for top partners has been recently proposed in Ref.~\cite{Buchkremer:2013bha}, where the vector-like quarks are embedded in different
representations of the weak $SU(2)$ group. We here consider a simplified model where the vector-like $T$-quark is an $SU(2)$ singlet, with couplings only to the third generation of SM quarks.
The benefit of using the simplified effective theory is that the results of the studies could be used to make predictions for more complex models including various types of top partners.
 The top partner sector of the model is described by the general effective Lagrangian~\cite{Buchkremer:2013bha}
\begin{eqnarray}
{\cal L}_{T} =&& \frac{gg^{\ast}}{2\sqrt{2}}[\bar{T}_{L}W_{\mu}^{+}
    \gamma^{\mu} b_{L}+\frac{g}{\sqrt{2}c_W}\bar{T}_{L} Z_{\mu} \gamma^{\mu} t_{L}- \frac{m_{T}}{\sqrt{2}m_{W}}\bar{T}_{R}ht_{L} -\frac{m_{t}}{\sqrt{2}m_{W}} \bar{T}_{L}ht_{R} ]+ h.c.,
  \label{TsingletVL}
\end{eqnarray}
where $m_T$ is the top partner mass and $g^{\ast}$ parametrizes the single production coupling in association with a $b$ or a top-quark. $g$ is the $SU(2)_L$ gauge coupling constant, $c_{W}=\cos \theta_W$ and $\theta_W$ is the Weinberg angle.

In this simplified model, the top partner couplings to
the SM particles are rather severely constrained by electroweak constraints as well as by
the direct measurement of $V_{tb}$~\cite{vtb}. However, such constraints
can be significantly altered in most realistic models including the vector-like quark with two or more partner multiplets~\cite{bound}.
Here we take a conservative range for the coupling parameter~\cite{atlas13,vtb}:
$g^{\ast}\leq 0.5$.

\section{Event generation and analysis}

During the simulation, we first extract the model file \cite{web} of the singlet vector-like top partners by using the \texttt{FeynRules} package \cite{feynrules}. The leading order cross sections are calculated using \texttt{MadGraph5-aMC$@$NLO} \cite{mg5} with CTEQ6L parton distribution function (PDF)~\cite{cteq} and the renormalization and factorization scales are set dynamically by default. The collider parameter is taken to be $E_e= 140$ GeV and $E_p=7$ TeV, corresponding to a c.m. energy of approximately $\sqrt{s}= 1.98$ TeV. The SM input parameters relevant in our study are taken from \cite{pdg}.

\begin{figure}[thb]
\begin{center}
\centerline{\epsfxsize=9cm \epsffile{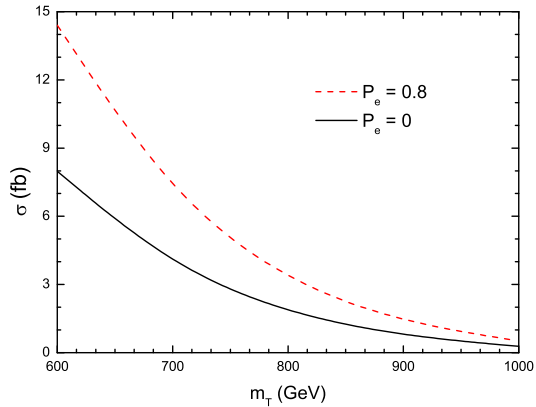}}
\caption{The dependence of the cross sections $\sigma$ on $m_T$ for $g^{\ast}=0.2$ and positron beam polarization $p_{e}=0, ~0.8$.  }
\label{cross}
\end{center}
\end{figure}

In Fig.~\ref{cross}, we show the single production cross sections of the vector-like $T$ depending on their masses at the LHeC for $g^{\ast}=0.2$ and positron beam polarization $p_{e}=0, 0.8$. We can see that the cross section for polarized can be about 1.8 times larger than that for unpolarized case. For $g^{\ast}=0.2$ and $p_{e}=0.8$, the cross sections are about 3.1, 1.3 and 0.5 fb for $m_T=800, 900$ and 1000 GeV, respectively.

\begin{figure}[htb]\vspace{-1.5cm}
\begin{center}
\centerline{\epsfxsize=16cm \epsffile{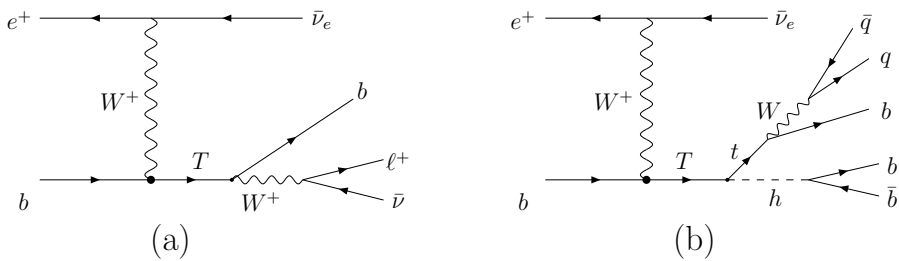}}
\vspace{-17.0cm}
\caption{The Feynman diagram for production of single $T$ quark including the decay chains $T\to bW\to b \ell \nu$ and $T\to th\to 2j+3b$.}
\label{epT}
\end{center}
\end{figure}

Next, we analyze the observation potential of each channel by performing a Monte
Carlo simulation of the signal and background events and applying the suitable selection cuts
to maximize the significance.
The Feynman diagram of
production and decay chain is presented in Fig.~\ref{epT}.

For the fixed $T$-quark mass, the corresponding free parameters are the coupling parameter $g^{\ast}$. We take three typical values of the $T$ quark mass: $m_T=800,900,1000$ GeV with $g^{\ast}=0.2$. We generate all event samples in this analysis at the leading order using  \texttt{MadGraph5-aMC$@$NLO}. Parton shower
(both initial and final) and hadronization effects have been dealt with by \texttt{PYTHIA6}~\cite{pythia}. For all the considered signals and backgrounds, the K-factors are taken to be 1~\cite{prd81-054018}.
We apply jet and lepton energy smearing according to the following energy resolution
formula~\cite{LHeC} $\delta E/E=a/\sqrt{E/\rm GeV}\oplus b$, where a is a sampling term and b is a constant term.
For jets we take $a = 45\%$ and $b = 3\%$, and for leptons we take $a = 8.5\%$ and $b = 0.3\%$. In our analysis, we assume a b-jet tagging efficiency of $\epsilon_{b} = 60\%$ and a corresponding
mistagging rate of $\epsilon_{g,u,d,s} = 1\%$ for light jets  and $\epsilon_{c} = 10\%$ for
a $c$-jet. Event analysis is performed by using \texttt{MadAnalysis5}~\cite{ma5}.
\subsection{The $T\to Wb$ channel}
In this section, we analyze the observation potential by performing a
Monte Carlo simulation of the signal and background events and explore the sensitivity of single top partner at the LHeC through the channel
\beq\label{signal}
e^{+}p \to T(\to bW^{+})\bar{\nu}_{e}\to bW^{+}( \to \ell^{+}\bar{\nu}_{\ell}) \bar{\nu}_{e}.
\eeq

For this channel, the typical signal is exactly one charged lepton, one $b$ jet and missing energy.
The main SM background are the processes containing a $W$ boson in the final state, such as
\beq
e^{+}p &&\to bW^{+}\bar{\nu}_{e}\to \ell^{+} + b + \slashed E_T^{miss}, \quad (\nu Wb)\nonumber \\
e^{+}p &&\to W^{+}(\to \ell^{+}\bar{\nu}_{\ell})j\bar{\nu}_{e}\to \ell^{+} + j + \slashed E_T^{miss}, \quad(\nu Wj)
\eeq
where one light jet might be faked as $b$ jet. For the $\nu Wb$ process, the dominant process is the single top production process in which the top quark decay in to $Wb$. We also checked that other background processes, such as the di-boson production are negligible with the selection cuts.

To reduce the backgrounds, we pick up the events that included exactly one isolated lepton and one b-tagged jet, then impose the following basic cuts:
\beq
p_{T}^{\ell,b} > 20 \rm ~GeV, |\eta_{\ell}|<2.5, |\eta_{b}|<5, \slashed E_T^{miss}>20 \rm ~GeV.
\eeq

\begin{figure}[htb]
\begin{center}
\centerline{\epsfxsize=6.5cm\epsffile{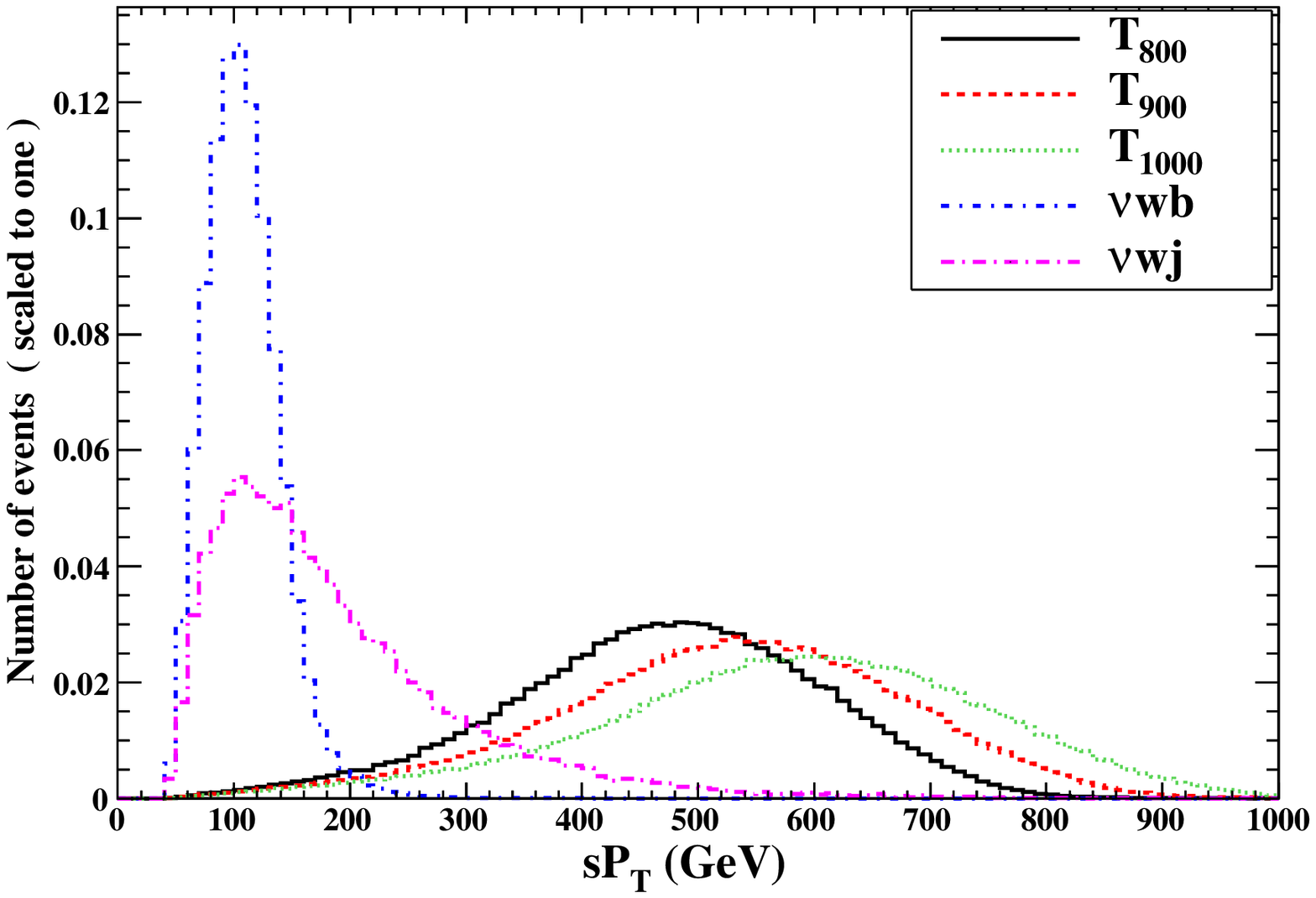}
\hspace{-0.5cm}\epsfxsize=6.5cm\epsffile{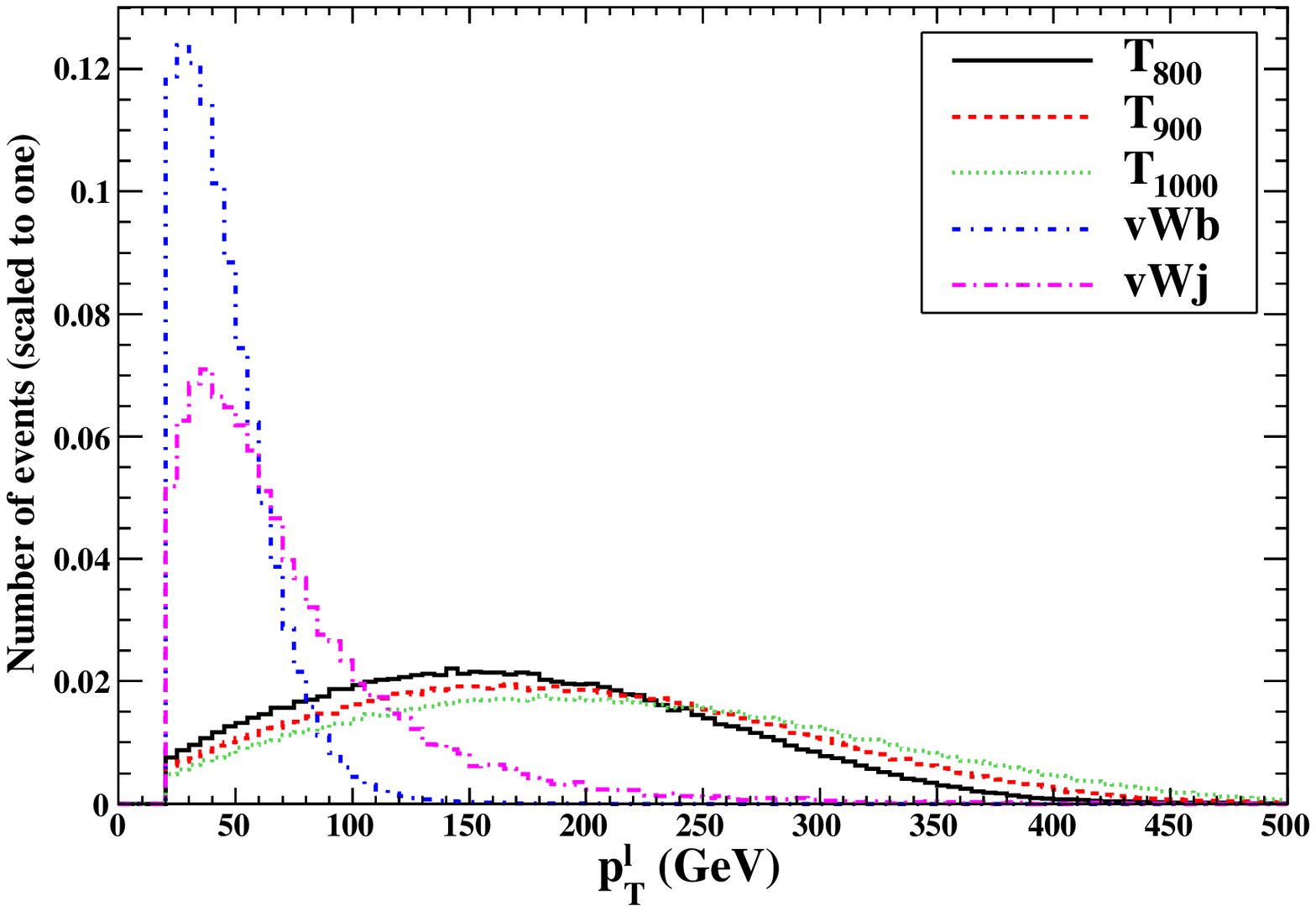}}
\centerline{\epsfxsize=6.5cm\epsffile{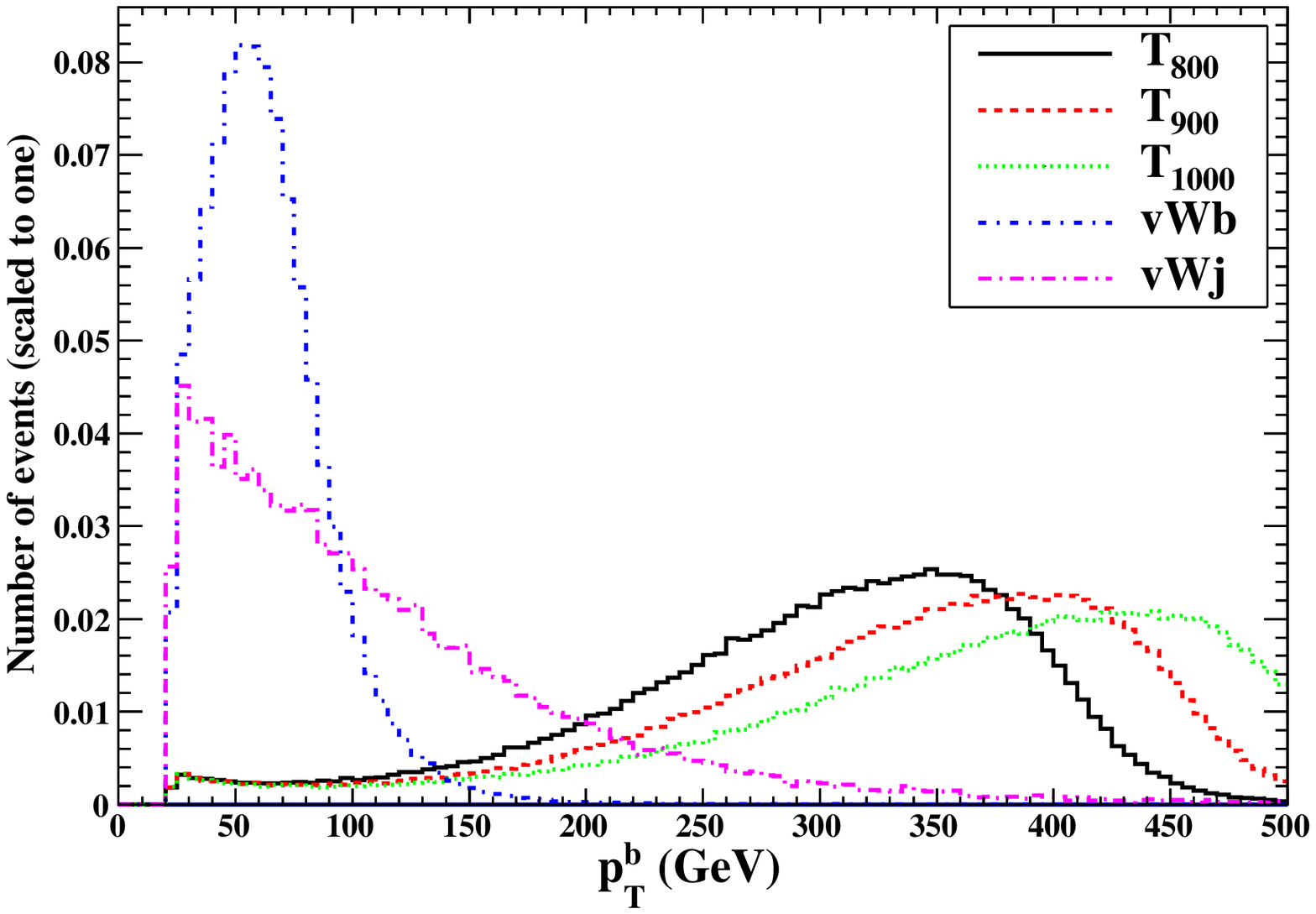}
\hspace{-0.5cm}\epsfxsize=6.5cm\epsffile{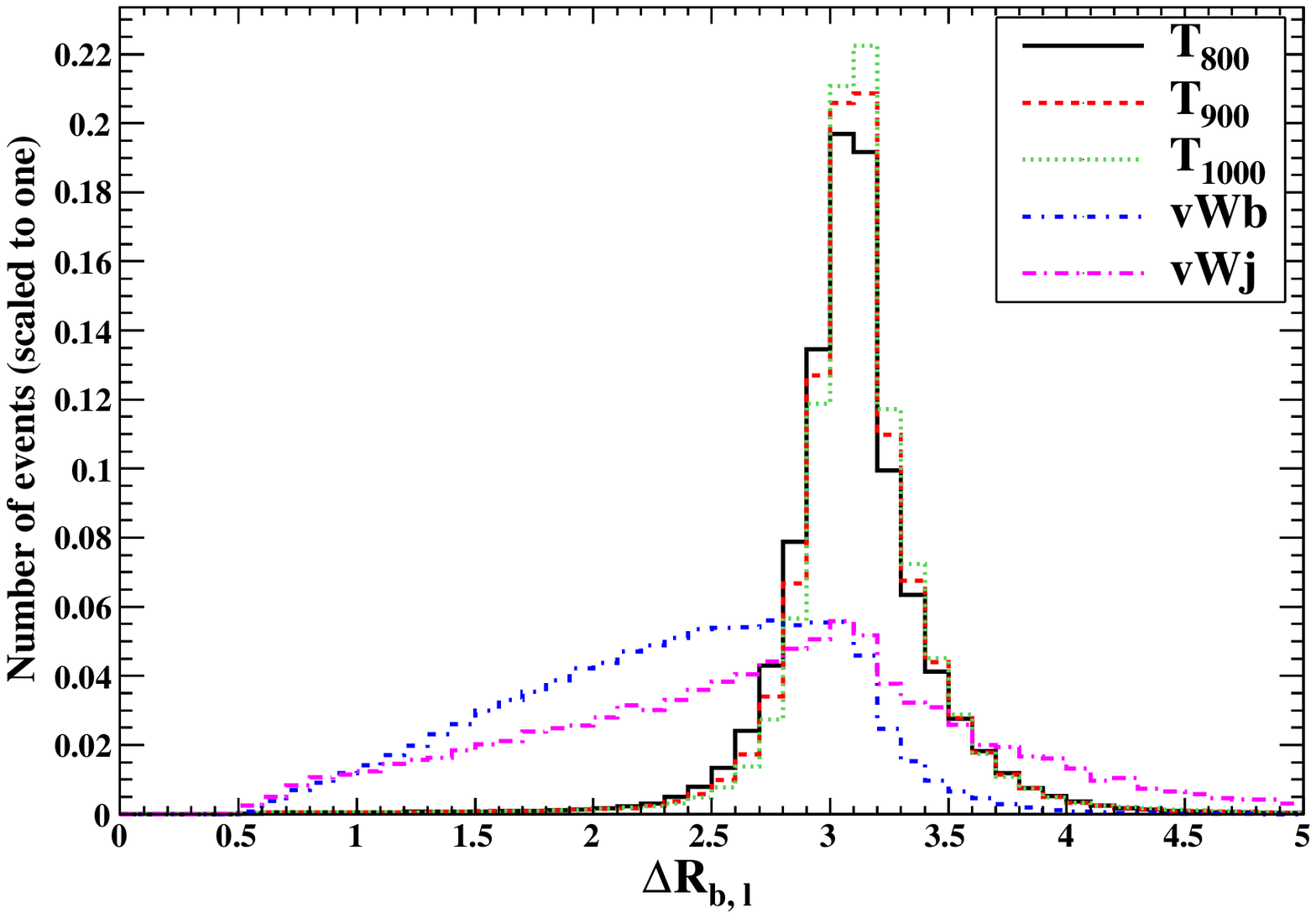}}
\caption{Normalized distributions of the scalar sum of the transverse momenta $\rm sP_{T}$, the transverse momentums ($p_{T}^{\ell}$ and $p_{T}^{b}$) and $\Delta R(b,\ell)$ for the signals and backgrounds.}
\label{spt}
\end{center}
\end{figure}

In Fig.~\ref{spt}, we show the normalized distributions of the scalar sum of the transverse momenta ($\rm sP_{T}$) of the $b$-tagged jet and the lepton, the transverse momentums  $p_{T}^{\ell, b}$ and the variable $\Delta R(b,\ell)$ for the signals and backgrounds. Here $\Delta R=\sqrt{(\Delta\phi)^{2}+(\Delta\eta)^{2}}$ is the particle separation with $\Delta\phi$ and $\Delta\eta$ being the separation in the azimuth angle and rapidity respectively. Based on these kinematical distributions, we impose the following cuts to
get a high significance:
\begin{itemize}
\item Cut 1: $\rm sP_T> 500 \rm ~GeV$.
\item Cut 2: $p_{T}^{\ell}> 100 \rm ~GeV$, $p_{T}^{b}> 250 \rm ~GeV$ and $2.8 < \Delta R(b,\ell)< 3.5$.
\end{itemize}

\begin{figure}[htb]
\begin{center}
\vspace{-0.5cm}
\centerline{\epsfxsize=9cm \epsffile{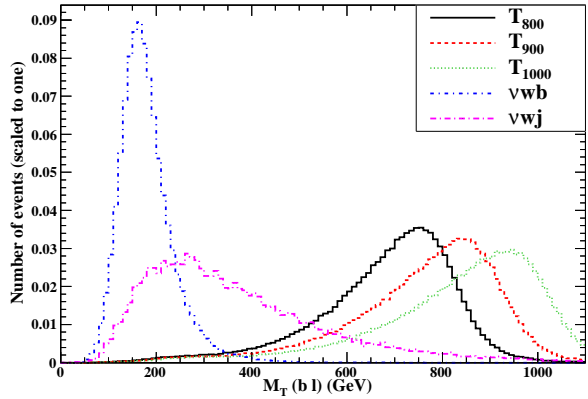}}
\caption{Normalized transverse mass distribution of the $b\ell \slashed E_T$  system for the signals and backgrounds.}
\label{mbl}
\end{center}
\end{figure}

In Fig.~\ref{mbl}, we show the transverse mass distribution for the $b\ell \slashed E_T$ system,
which has been defined in Ref.~\cite{epjc-74-3103}.
From this figure, we can see that the distributions of signal have peaks
around the $T$-quark mass while background distributions
turn over at lower masses. Thus we can further reduce the backgrounds by the following cut:
\begin{itemize}
\item Cut 3: $M_T>700 \rm~GeV$.
\end{itemize}

\begin{table}[htb]
\begin{center}
\caption{The cut flow of the cross sections (in fb) for the signal and backgrounds
at the LHeC with $E_e= 140$ GeV and $E_p=7$ TeV. Here we take $g^{\ast}=0.2$. \label{cutflow1}}
\vspace{0.5cm}
\begin{tabular}{c|c|c|c|c  c|c}
\hline
\multirow{2}{*}{Cuts}& \multicolumn{3}{c|}{signal}&\multicolumn{3}{c}{backgrounds}\\ \cline{2-7}
 &{800~GeV} &{900~GeV}&{1000~GeV}&$\nu Wb$&$\nu Wj$&total \\  \cline{1-7}
\hline
Basic cuts& 0.3&0.13&0.05&768&34&802\\ \hline
Cut 1&0.13&0.077&0.036&0.006&0.61&0.62\\ \hline
Cut 2&0.12&0.07&0.032&0.006&0.31&0.32\\ \hline
Cut 3&0.1&0.064&0.031&0.003&0.23&0.23\\ \hline
\end{tabular} \end{center}\end{table}

We present the cross sections of the signal and backgrounds after imposing
the cuts in Table~\ref{cutflow1}.
From the numerical results,
one can see that the backgrounds are suppressed very efficiently after imposing the selections. For example, the production cross section of the $\nu Wb$ background drops from 768 fb to 0.006 fb, with a reject efficiency more than $99\%$.
The dominant SM background come from the $\nu Wj$ production process and is about 0.23 fb after imposing the selections.

\subsection{The $T\to th$ channel}
Next, we analyze the observation potential and explore the sensitivity of single $T$-quark at the LHeC through the channel
\beq\label{signal2}
e^{+}p \to T(\to th)\bar{\nu}_{e}\to t( \to W^{+}b\to jj'b)h(\to b\bar{b}) \bar{\nu}_{e}.
\eeq

The main SM background are the processes:
\beq
e^{+}p &&\to th\bar{\nu}_{e}\to t( \to W^{+}b\to jj'b)h(\to b\bar{b}) + \slashed E_T^{miss}, \quad (\nu th)\nonumber \\
e^{+}p &&\to tZ\bar{\nu}_{e}\to t( \to W^{+}b\to jj'b)Z(\to b\bar{b}) + \slashed E_T^{miss}. \quad(\nu tZ)
\eeq
Besides, the single top and single $W$ plus jets production processes are also the backgrounds, where one light jet might be faked as $b$ jet.

The event selection in the 3b-tagging case first requires at least five jets satisfying the following basic cuts:
\begin{itemize}
\item
$p_{T}^{j} > 20 \rm ~GeV$, $|\eta_{j}|<5$, $\Delta R_{jj} > 0.4$.
\item
There are at least three b-tagged jets with $|\eta_{b}|<5$.
\item
Events with additional charged leptons are vetoed.
\end{itemize}
\begin{figure}[htb]
\begin{center}
\centerline{\epsfxsize=10cm\epsffile{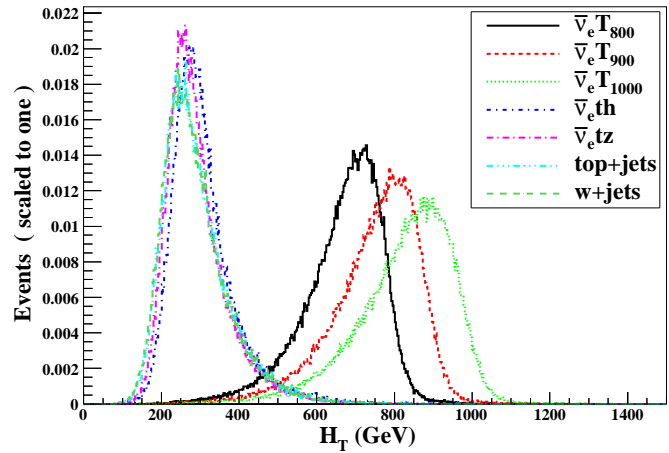}}
\caption{Normalized distributions of the $H_T$ for the signals and backgrounds.}
\label{th}
\end{center}
\end{figure}
\begin{figure}[htb]
\begin{center}
\centerline{\epsfxsize=10cm\epsffile{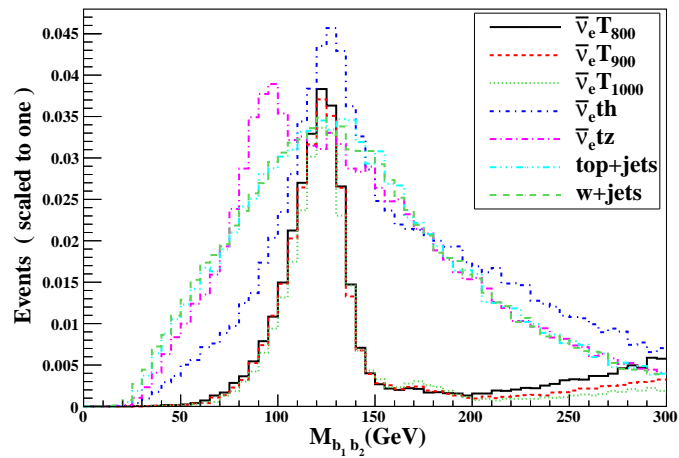}}
\caption{Normalized distributions of the $m_h$ for the signals and backgrounds.}
\label{mh}
\end{center}
\end{figure}

In Fig.~\ref{th}, we show the $H_T$ distribution for the different considered processes, where $H_T$ denotes
the scalar sum of the transverse momenta of the final state jets. The
signal distribution has a considerable tail for larger values of $H_T$ compared to background events. Therefore we choose the $H_T$ cut
\begin{itemize}
\item Cut 1: $H_T>600 \rm~GeV$.
\end{itemize}

Fig.~\ref{mh} illustrates the distribution of the reconstructed Higgs boson mass of the signal
and backgrounds. In order to suppress the $tZ\bar{\nu}_{e}$, $tbb$ and $Wbbj$ backgrounds, we require the mass of the Higgs boson to
satisfy
\begin{itemize}
\item Cut 2: $|m_{bb}-m_h|<20 \rm~GeV$.
\end{itemize}

We list the results after imposing various kinematic cuts in Table~\ref{cutflow2}.
From the numerical results,
one can see that all the backgrounds are also suppressed efficiently after imposing these selections.
However, due to the small production cross section for the signals, the large value of the coupling parameter $g^{\ast}$ and the high integrated luminosity are needed to produce more final events.
Note that here our results are conservative, the analysis presented can be further improved
in several aspects. First is of course a more realistic
estimation of the signal and backgrounds including
parton shower and more detailed detector effects.
Secondly, we have only applied a cut-based analysis with very
simple variables. A further multivariate analysis may
deliver additional gain in sensitivity. Furthermore, other techniques such as Boosted Decision Trees(BDT), Heidelberg-Eugene-Paris top-tagger (HEPtopTagger) and jet dipolarity~\cite{BDRS,top-tagger,Hook-jhep} may be more useful to enhance the significance.

\begin{table}[htb]
\begin{center}
\caption{The cut flow of the cross sections (in $10^{-2}$ fb) for the signal and backgrounds
at the LHeC with $E_e= 140$ GeV and $E_p=7$ TeV. Here we take $g^{\ast}=0.2$. \label{cutflow2}}
\vspace{0.5cm}
\begin{tabular}{c|c|c|c|c c c c |c}
\hline
\multirow{2}{*}{Cuts}& \multicolumn{3}{c|}{signal}&\multicolumn{5}{c}{backgrounds} \\ \cline{2-9}
 &{800~GeV} &{900~GeV}&{1000~GeV}&$\bar{\nu}_{e}th$&$\bar{\nu}_{e}tZ$&$t+2b$&$W+3b$&total \\  \cline{1-9}
\hline
Basic cuts&5.2&1.92&0.64&9.5&23&27&28&87.5\\ \hline
Cut 1&4.4&1.84&0.6&0.16&0.39&0.43&0.47&1.45\\ \hline
Cut 2&1.0&0.4&0.12&0.04&0.026&0.03&0.03&0.13\\ \hline
\end{tabular} \end{center}\end{table}

\begin{figure}[htb]
\begin{center}
\centerline{\epsfxsize=8cm \epsffile{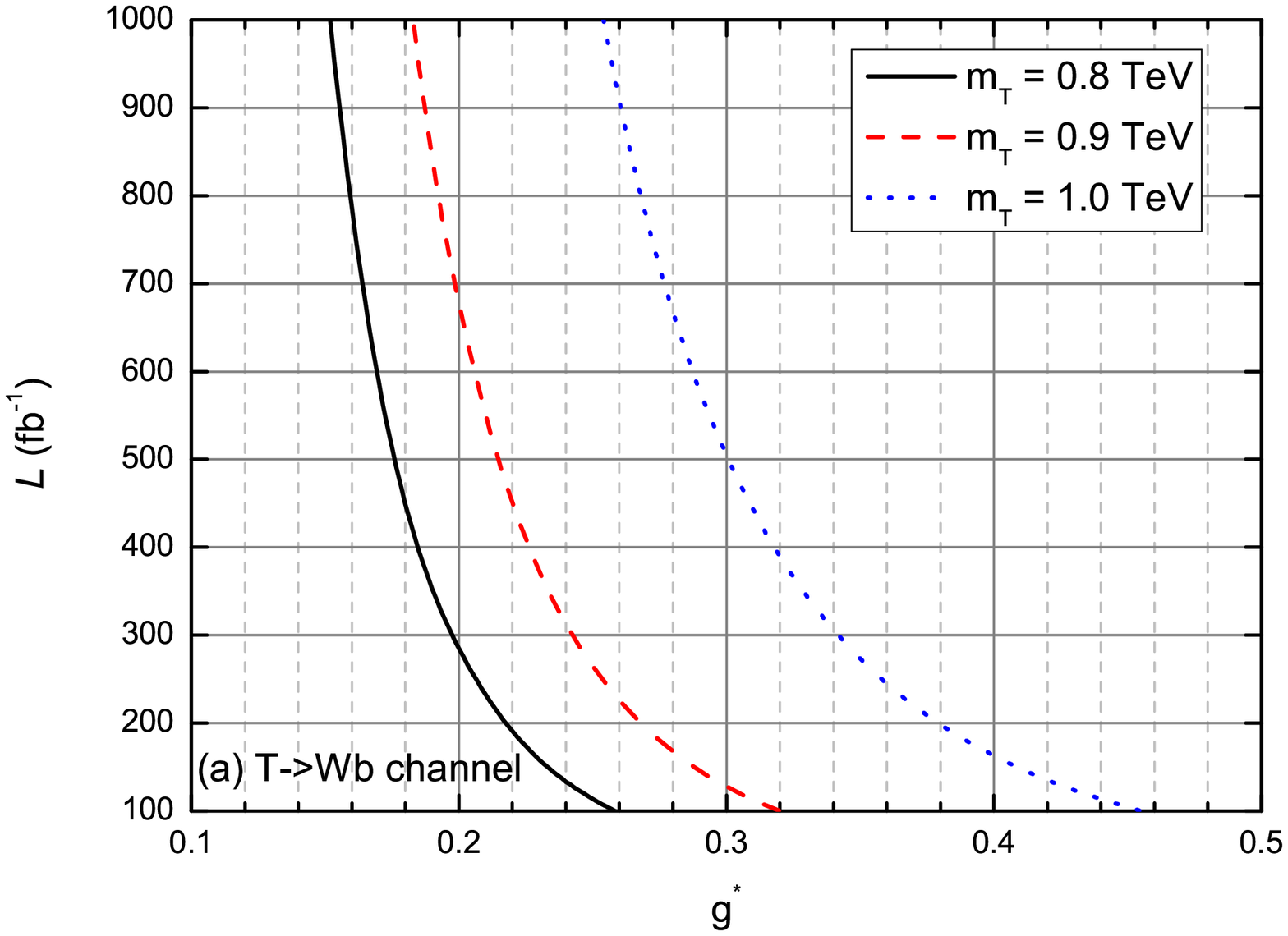}\epsfxsize=8cm \epsffile{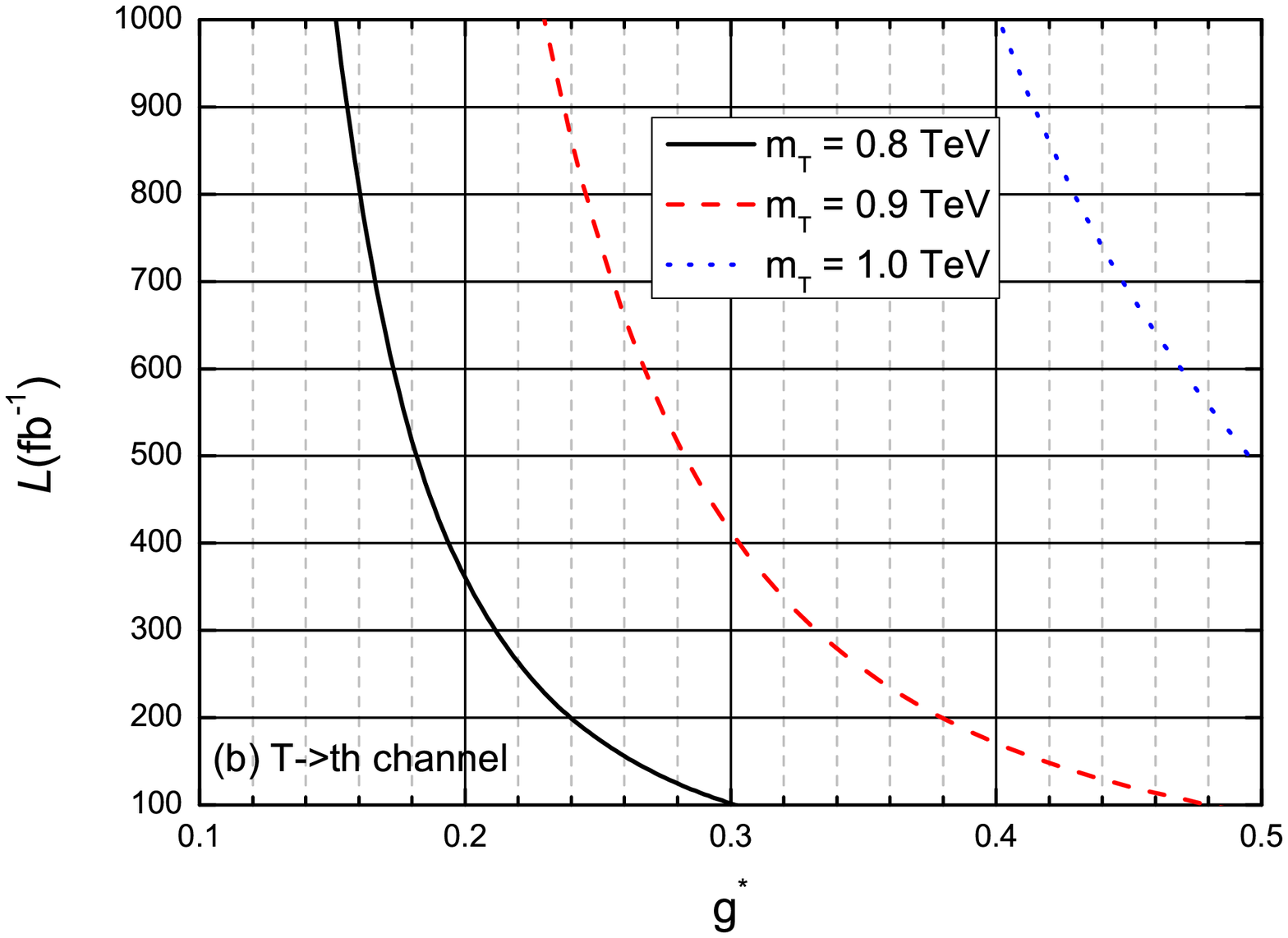}}
\caption{$3\sigma$ contour plots for the signal in $L-g^{\ast}$  at the LHeC for (a) $T\to Wb$ channel, and (b) $T\to th$ channel. }
\label{ss}
\end{center}
\end{figure}

To estimate the observability quantitatively, we adopt the significance measurement~\cite{ss}:
$SS=\sqrt{2L[(\sigma_S+\sigma_B)\ln(1+\sigma_S/\sigma_B)-\sigma_S]}$,
where $L$ is the integrated luminosity, $\sigma_S$ and $\sigma_B$ are the cross sections of signal and background, respectively. In Fig.~\ref{ss}, we show the excluded $3\sigma$ reaches in the plane of the integrated luminosity and the coupling parameter $g^{\ast}$ for two channels. For $m_{T}= 800$,~900 and 1000 GeV, we can see that the upper limits for the $T\to Wb$ channel on the size of $g^{\ast}$ are respectively given as about 0.2, 0.24 and 0.34 for $L= 300$ fb$^{-1}$, and changed as about 0.15, 0.18 and 0.25 when the integrated luminosity is 1000 fb$^{-1}$. For the $T\to th$ channel, the high integrated luminosity is needed to enhance the production events due to the small production rates. For the same three top partner masses, we can see that the upper limits on the size of $g^{\ast}$ are respectively given as about 0.15, 0.23 and 0.4 when the integrated luminosity is 1000 fb$^{-1}$.

\section{conclusion and discussion}
The new heavy vector-like top partners with charge $2/3$ appear in many new physics models beyond the SM. In order to
be as model-independent as possible we exploited a simplified model with only two free
parameters, the heavy top mass and an electroweak coupling constant. In this paper we described the future LHeC potential to search for the heavy vector-like $T$-quark via the $T\to bW^{+}$ and $T\to th$ decay modes. We investigated the observability of the heavy vector-like top partner $T$ production through the processes $e^{+}p \to T(\to bW^{+})\bar{\nu}_{e}\to bW^{+}( \to \ell^{+}\bar{\nu}_{\ell}) \bar{\nu}_{e}$ and $e^{+}p \to T(\to th)\bar{\nu}_{e}\to t( \to jj'b)h(\to b\bar{b}) \bar{\nu}_{e}$ at the LHeC with $E_e= 140$ GeV (with 0.8 polarization) and $E_p=7$ TeV. Since the single $T$ production depends on the $TWb$ coupling and the $T$-quark mass, we constrain the parameter space in the plane of the integrated luminosity and the coupling parameter $g^{\ast}$. In the $bW$ channel, we rely on the large $p_T$ of the lepton, the $b$-jet and large missing energy to suppress the backgrounds. In the $th$ channel, we consider the all-hadronic final states and a large $H_T$ cut can efficiently suppress the backgrounds. For $m_{T}$= 800,~900 and 1000 GeV, the upper limits on the size of $g^{\ast}$ are respectively given as about 0.15, 0.18 and 0.25 with the high luminosity of 1000 fb$^{-1}$. We expect our analysis can represent a complementary candidate to pursue the search of a possible singlet top partner below $\sim 1$ TeV.

In some typical models such as the minimal Composite Higgs (CH) models~\cite{jhep1304-004}
 and the littlest Higgs model with T-parity (LHT)~\cite{lht}, the heavy
$T$-quark couplings only to the third generation of SM quarks. Our results can
be straightforwardly mapped within the context of the CH model and the LHT
model, namely with
\beq
g^{\ast}&\simeq& \frac{y m_{W}}{gm_{T}},\quad (\rm CH) \nonumber \\
g^{\ast}&\simeq&\frac{R^{2}}{1+R^{2}}\frac{v}{f}+{\cal O}(\frac{v^{2}}{f^{2}}), \quad (\rm LHT)
\eeq
where $y$, $R$ and $f$ are the model parameters (for more detail, see e.g.\cite{jhep1304-004,lht}).
For the low mass reign (i.e., $m_T\lsim 1$ TeV), we expect our analysis can represent a
viable and complementary candidate to pursue the search of the singlet vector-like
$T$-quarks.

Now we compare the discovery reach in our results with other phenomenological studies at the LHC.
In a simplified composite Higgs model, the authors of Ref.~\cite{NP-shu} project at $\sqrt{s}=8$ TeV and 25 fb$^{-1}$ of integrated luminosity the exclusion potential with $Br(T\to Wb)=0.5$, obtaining a
strong constraint on the $TWb$ coupling ($g_{TbW}/g^{SM}_{TbW}<0.2\sim0.3$) for $m_T\subset (700,1000)$ GeV.  For the $T\to tZ$ channel, the authors of Ref.~\cite{jhep1501-088} studied the dilepton signals  at $\sqrt{s}=13$ TeV with 300 fb$^{-1}$ of integrated luminosity and the results show that the signal from the $T\to tZ\to (qq'b)(\ell^{+}\ell^{-})$ channel
is within the range of possible evidence for top partner masses up to roughly 1450 GeV with $g^{\ast}\lsim 0.5$, while being still sensitive to $g^{\ast}$ couplings down to 0.05 for $m_T=0.8$ TeV. Analogously,
 the authors of Ref.~\cite{jhep2015-02-032} studied the trilepton signals from the $T\to tZ\to (b\ell\nu)(\ell^{+}\ell^{-})$ channel at the LHC with $\sqrt{s}=13$ TeV, the results show that the trilepton signal can probe possible top partner masses up to roughly 1700 GeV with
$g^{\ast}\lsim 0.5$, while being still sensitive to $g^{\ast}$ couplings down to 0.1 for $m_T=0.8$ TeV. Thus we can see that, for the typical mass parameter $m_T=0.8$ TeV, the LHeC sensitivity for the $TbW$ coupling strength is smaller than the sensitivity limits of LHC as $g^{\ast}\lsim 0.1 (0.05)$ with an integrated luminosity of 300 fb$^{-1}$ at $\sqrt{s}=13$ TeV~\cite{jhep1501-088,jhep2015-02-032}.
In our previous study~\cite{plb-liu}, the decay of $T\to t(\to b j j')Z( \to \ell^{+}\ell^{-})$ channel from the single top partner produced at the LHeC are investigated by considering the mixing between the top partner with the first and the third generation quarks. For a high integrated luminosity of 1 ab$^{-1}$, the upper limits on the size of $g^{\ast}$ are given as $g^{\ast}\leq 0.29$ for $R_L=0$ (only couplings with the third generation SM quarks) and $m_T=0.8$ TeV.  Thus we expect our analysis can represent a viable and complementary candidate to pursue the search of a possible vector-like top partner with low masses (i.e., $m_T\lsim 1$ TeV).

\begin{acknowledgments}
This work is supported by the Joint Funds
of the National Natural Science Foundation of China (Grant No. U1304112), the Foundation of He¡¯nan Educational Committee (Grant No. 2015GGJS-059) and the Foundation of Henan Institute of Science and Technology (Grant No. 2016ZD01).
\end{acknowledgments}


\end{document}